\documentclass{emulateapj}

\usepackage{graphicx}
\usepackage{epsfig}
\usepackage{amssymb}
\usepackage{amsmath}
\usepackage{graphicx}
\usepackage{ifpdf}
\usepackage{natbib}
\usepackage{bm}
\usepackage{xcolor}
\usepackage{comment}
\usepackage{mnsymbol}
\usepackage[export]{adjustbox}
\usepackage{lineno}

\definecolor{darkgreen}{rgb}{0.0,0.5,0.0}
\usepackage[colorlinks,citecolor=darkgreen]{hyperref}

\newcommand{\sgn}{\operatorname{sgn}}

\begin{document}

\title{On the escape of low-frequency waves from magnetospheres of neutron stars}
\shortauthors{Golbraikh and Lyubarsky}
\author{Ephim Golbraikh}
\author{Yuri Lyubarsky}
\affil{Physics Department, Ben-Gurion University of the Negev, POB 653, Be'er-Sheva 84105, Israel}
	
\date{\today}

\begin{abstract}
We study the nonlinear decay of the fast magnetosonic into the Alfv\'en waves in relativistic force-free magnetohydrodynamics. The work has been motivated by models of pulsar radio emission and fast radio bursts (FRBs), in which the emission is generated in neutron star magnetospheres at conditions when not only the  Larmor but also the plasma frequencies significantly exceed the radiation frequency. The decay process places limits on the source luminosity in these models. We  estimated the decay rate and showed that the phase volume of  Alfv\'en waves available for the decay of an fms wave is infinite. Therefore the energy of fms waves  could be completely transferred to the small-scale Alfv\'en waves not via a cascade, as in the Kolmogorov turbulence, but directly. Our results explain the anomalously low radio efficiency of the Crab pulsar and 
show that FRBs could not be produced well within magnetar magnetospheres.

\end{abstract}

\date{Accepted ---. Received ---; in original ---}
\label{firstpage}

\keywords{magnetohydrodynamics -- plasma astrophysics -- radiative processes -- radio transient sources -- pulsars}

\maketitle

\section{Introduction}

Nonlinear effects play an important role in powerful compact sources of radio emission, such as pulsars and fast radio bursts (FRBs). For example, in nonmagnetized plasmas, the induced Compton and Raman scattering could even prevent the escape of the waves. In the strongly magnetized magnetosphere of neutron stars, the electromagnetic waves propagate in two orthogonally polarized modes:  the so-called O-mode is polarized in the plane set by the background magnetic field and the propagation direction, whereas the X-mode is polarized perpendicularly to the magnetic field and the propagation direction. Only the O-mode is subject to induced scattering because the electric field of this mode has a component along the background magnetic field. In the field of the X-mode, the particles oscillate only due to the weak $\mathbf{E\times B}$ drift; therefore, the scattering is suppressed. 

The O-mode could propagate only if, in the rest frame of the plasma, the wave's frequency exceeds the plasma frequency.  
If the density of the plasma is high enough so that not only the Larmor but also the plasma frequency is well above the wave frequency, the two magnetohydrodynamic (MHD) waves could propagate: the fast magnetosonic (fms) and the Alfv\'en waves. The fms wave is polarized perpendicularly to the background magnetic field and the propagation direction. When this wave propagates towards decreasing plasma density, it is smoothly converted into the X-mode and could escape from the system. The Alfv\'en wave does not escape; it follows the curved magnetic field lines and eventually decays via the Landau damping \citep{Arons_Barnard86}.

Thus, in dense magnetospheres (such that both the Larmor and the plasma frequencies are above the emission frequency), the radiation propagates in the form of fms waves, independently of the emission mechanism. In this  case, the nonlinear decay of fms into Alfv\'en waves could strongly affect the outgoing radiation. The goal of this paper is to study the decay process. 
In the magnetospheres of neutron stars, the magnetic energy significantly exceeds the plasma energy; therefore, the wave interaction could be considered in the scope of relativistic force-free MHD. The wave energy is well below the energy of the background field; therefore, we could employ the methods of the weak turbulence theory (see, e.g., \citealt{Zakharov_etal92}). Namely, we write down and solve the kinetic equation for the waves. 
Note that in the non-relativistic case, solutions to the kinetic equations for fms and Alfv\'en waves were investigated
both analytically \citep{Kuznetsov01} and numerically \citep{Chandran05,Chandran08}.

The paper is organized as follows. In sect.\ 2, we write down the kinetic equations for MHD waves in the relativistic force-free regime. In sect.\ 3, we analyze the equations, estimate the nonlinear decay rate of fms waves,  and qualitatively describe the kinetics of the decay of fms into Alfv\'en waves.  In sect.\ 4, we solve the kinetic equations numerically, confirming our qualitative analysis. The 
implications of our findings for FRBs are outlined in sect.\ 5. Conclusions are presented in sect.\ 6.

\section{Nonlinear interaction of MHD waves in force-free regime}

In this paper, we address weakly nonlinear interactions of MHD waves in the relativistic force-free regime when the plasma energy density, including the rest mass energy, is negligible compared to the magnetic energy density. This implies that the widely used magnetization parameter, $\sigma=B^2/4\pi\rho c^2$, is infinite. Here $B$ is the background magnetic field, and $\rho$ is the plasma density.     

In the force-free limit, there are two MHD waves: the fms wave, polarized perpendicularly to the background magnetic field and to the propagation direction, and the Alfv\'en wave, polarized perpendicularly to the background magnetic field in the plane set by the field and the propagation direction. The dispersion equations in the force-free limit are very simple:  
\begin{equation}
    \omega=ck
\label{fms_disp}\end{equation}
for the fms  waves and  \begin{equation}
    \omega=ck\vert\cos\theta\vert
\label{ALfven_disp}\end{equation}
for the Alfv\'en waves.
Here $\omega$ is the frequency, $k$ the wave vector, and $\theta$ the angle between the wave vector and the background magnetic field. 

The interaction of weakly nonlinear MHD waves in the force-free limit was studied by \citet{Thompson_Blaes98} and \citet{ Lyubarsky19}. The strongest is the interaction of three waves satisfying the resonance conditions
\begin{equation}
\omega=\omega_1+\omega_2;\quad \mathbf{k}=\mathbf{k}_1+\mathbf{k}_2,
\label{conservation}\end{equation}
which in fact represent energy and momentum conservation.
It follows from the dispersion relations that the conservation laws are satisfied for the decay of an fms wave into an fms and an Alf\'ven waves ($S\leftrightarrow S+A$) and into two Alf\'ven waves ($S\leftrightarrow A+A$). Of course, reverse merging processes are also possible. 
The conservation laws also permit the process $S\leftrightarrow S+S$ for the aligned fms waves, but in the force-free limit, the probability of the process is zero \citep{Lyubarsky19}. At a finite $\sigma$, the weakly nonlinear interaction of aligned fms waves becomes possible; it leads to the steepening of the waves and formation of shocks \citep{Levinson_vanPutten97,Lyubarsky03}.   

The three-wave interaction of Alfv\'en waves, $A+A\rightarrow A$, is possible only if one of two waves has zero frequency. Then an anisotropic Alfv\'en cascade develops, transferring the energy to waves with large components of the wave vector perpendicular to the background field, $k_\perp\gg k_\|$
\citep{Montgomery_Matthaeus05,Ng_Bhattacharjee96, Goldreich_Sridhar97}.  The important point is that  the zero-frequency Alfv\'en waves cannot be treated as linear waves.  
 Such interaction occurs when  
the field lines of the background magnetic field wander away. 
The Alfv\'en waves are stretched when propagating along diverging field lines, so the wave vector component perpendicular to the background field increases. Thus a cascade is formed, redistributing the Alfv\'en waves towards the high-$k_\perp$ domain, where they eventually decay. 
\citet{Goldreich_Sridhar97} presented a qualitative explanation of how small turbulent fluctuations, $\delta B\ll B$, could lead to divergent field lines. Assume that the mean magnetic field is directed along $z$-axis and describe the turbulence as an ensemble of localized wave packets with the longitudinal and transverse scales $l_\|$ and $l_\perp$, correspondingly, and the amplitude $\delta B$. The local magnetic field line turns within a wave packet by the angle $\theta\sim\delta B/B$, so the field line deviates from the initial position by $\theta l_\|$.  Adding random deviations, one finds that the average displacement of the field line grows with the distance, $s\sim\theta\sqrt{l_\|z}$. Therefore, the field lines that were initially separated by $l_\perp$ diverge. This picture implicitly assumes a long wavelength tail in the fluctuation spectrum. Namely, the amplitude of the fluctuations at the scale $z$ is $\Delta B\sim Bs/z\sim \delta B\sqrt{l_\|/z}$. This implies  $\Delta B^2
\propto k_\|$ so that the spectral power of the turbulence, $\Delta B^2_{k_\|}=d\Delta B^2/dk_\|$, goes to a constant at $k_\|\to 0$.  Such a spectrum is obtained if the turbulence is presented as an ensemble of bell-shaped fluctuations. We here deal with Alfv\'en waves produced by the decay of fms waves. The fms  waves with the frequency $\omega$ produce Alfv\'en waves with the wave vector $k_\|\sim\omega/c$. The production rate is proportional to $\omega$ and the energy density of fms waves at this frequency, $U$ (see eq.\ \ref{q0}). Both these quantities decrease towards smaller frequencies, so the spectrum of the produced waves is cut off at long wavelengths. Therefore, in the case of interest, the three-wave Alfv\'en cascade is suppressed and could be neglected.

 Of course, the Alfv\'en cascade could develop via four-wave interactions, $A+A\to A+A$ \citep{Sridhar_Goldreich94,Goldreich_Sridhar95}. However, we assume that the turbulence is weak, i.e., the system's evolution is governed by the lowest-order processes.  At $k_\perp\sim k_\|$, the rate of four-wave processes is lower than that of three-wave processes by the ratio of the wave energy to the energy of the background field, $8\pi U/B_0^2$.   
We will show that the fms waves with the frequency $\omega$ decay first into Alfv\'en waves with $k_\perp\sim k_\|\sim\omega/c$, so if the above ratio is small, the fms-to-Alfv\'en transformation initially occurs in the weak turbulence regime. However, higher $k_\perp$ Alfv\'en waves are produced in the course of time. It is well known that the role of nonlinearity grows at higher $k_\perp$ so that turbulence ceases to be weak at $k_\perp/k_\|\sim c/\delta v$, where $\delta v=c\delta B/B_0$ is the velocity of turbulent motions, $\delta B$ the fluctuating magnetic field \citep{Goldreich_Sridhar95}.
However, we will see that the spectrum of Alfv\'en waves reaches high $k_\perp$ only when the fms energy decreases $\sim (ck_\perp/\omega)^2$ times.  
Therefore, most of the transformation process occurs in the weak turbulence regime.

In astrophysical applications, we typically deal with wide spectra and random phases of waves. Then the wave field is conveniently described in the quantum language via the occupation numbers, $n_{\mathbf{k}}$, which are related to 
the wave energy density of a mode with the wave vector $\mathbf{k}$
as $E_{\mathbf{k}}=\omega_{\mathbf{k}}n_{\mathbf{k}}$.
Since we deal with two types of waves, we denote the occupation numbers of Alfv\'en waves by $n_{\mathbf{k}}$ and of fms waves by $N_{\mathbf{k}}$.
The evolution of the system is described by the kinetic equations for the waves (
e.g., \citealt{Zakharov_etal92}). For fms waves, these equations may be written as
\begin{equation}
\frac{\partial N_\mathbf{k}}{\partial t}=\sum_{\mathbf{k}_1,\mathbf{k}_2}\left[-R^{S\leftrightarrow A+S}_{\mathbf{k},\mathbf{k}_1,\mathbf{k}_2}+R^{S\leftrightarrow A+S}_{\mathbf{k}_2,\mathbf{k}_1,\mathbf{k}}-\frac 12R^{S\leftrightarrow A+A}_{\mathbf{k},\mathbf{k}_1,\mathbf{k}_2}\right],
\label{kinetic_fms}\end{equation}
where $R^{S\leftrightarrow S+A}_{\mathbf{k},\mathbf{k}_1,\mathbf{k}_2}$ and $R^{S\leftrightarrow A+A}_{\mathbf{k},\mathbf{k}_1,\mathbf{k}_2}$ are 
the rates of the $S\leftrightarrow S+A$ and $S\leftrightarrow A+A$ processes for the given set of wave vectors $\mathbf{k},\,\mathbf{k}_1,\,\mathbf{k}_2$.
Here, the first term describes the decay of the wave $\mathbf{k}$ into an fms and an Alfv\'en waves and the reverse process; the second is for the production of the fms $\mathbf{k}$-wave via decay of an fms wave with the frequency $\omega_1>\omega$ and the reverse process, and the third term is for the decay of the fms $\mathbf{k}$-waves into two Alfv\'en waves and the reverse process.  The factor $1/2$ in the third term takes into account double counting in the case of the decay into two waves of the same type. Similarly, the kinetic equation for the Alfv\'en waves is written as
\begin{equation}
\frac{\partial n_\mathbf{k}}{\partial t}=\sum_{\mathbf{k}_1,\mathbf{k}_2}\left[R^{S\leftrightarrow A+S}_{\mathbf{k}_2,\mathbf{k},\mathbf{k}_1}+R^{S\leftrightarrow A+A}_{\mathbf{k}_2,\mathbf{k}_1,\mathbf{k}}\right].
\label{kinetic_Alfven}\end{equation}
Now the factor $1/2$ does not appear in the corresponding term because one of the two Alfv\'en quanta is fixed.

Denoting the probabilities of the spontaneous decay processes as $W^{S\rightarrow A+S}_{\mathbf{k},\mathbf{k}_1,\mathbf{k}_2}$ and $W^{S\rightarrow A+A}_{\mathbf{k},\mathbf{k}_1,\mathbf{k}_2}$, correspondingly, taking into account the induced processes and using the detailed balance principle, one can write 
 \begin{gather}
R^{S\rightarrow A+S}_{\mathbf{k},\mathbf{k}_1,\mathbf{k}_2}=W^{S\rightarrow A+S}_{\mathbf{k},\mathbf{k}_1,\mathbf{k}_2}\left[N_\mathbf{k}(n_{\mathbf{k}_1}+1)(N_{\mathbf{k}_2}+1)\right.\\
\left.-(N_\mathbf{k}+1)n_{\mathbf{k}_1}N_{\mathbf{k}_2}\,\right]\delta(\mathbf{k}-\mathbf{k}_1-\mathbf{k}_2)\delta(\omega_\mathbf{k}-\omega_{\mathbf{k}_1}-\omega_{\mathbf{k}_2});\nonumber\\
R^{S\leftrightarrow A+A}_{\mathbf{k},\mathbf{k}_1,\mathbf{k}_2}=W^{S\rightarrow A+A}_{\mathbf{k},\mathbf{k}_1,\mathbf{k}_2}\left[N_\mathbf{k}(n_{\mathbf{k}_1}+1)(n_{\mathbf{k}_2}+1)\right.\\
\left.-(N_\mathbf{k}+1)n_{\mathbf{k}_1}n_{\mathbf{k}_2}\,\right]\delta(\mathbf{k}-\mathbf{k}_1-\mathbf{k}_2)\delta(\omega_\mathbf{k}-\omega_{\mathbf{k}_1}-\omega_{\mathbf{k}_2}).
\nonumber \end{gather}
In all cases of interest, $N_\mathbf{k}\gg 1$, therefore we neglect the linear in $N_\mathbf{k}$ terms, which describe spontaneous processes, and get
 \begin{gather}
R^{S\leftrightarrow A+S}_{\mathbf{k},\mathbf{k}_1,\mathbf{k}_2}=W^{S\rightarrow A+S}_{\mathbf{k},\mathbf{k}_1,\mathbf{k}_2}\left(N_\mathbf{k}n_{\mathbf{k}_1}+N_\mathbf{k}N_{\mathbf{k}_2}\right.\\\left.-n_{\mathbf{k}_1}N_{\mathbf{k}_2}\,\right)\delta(\mathbf{k}-\mathbf{k}_1-\mathbf{k}_2)\delta(\omega_\mathbf{k}-\omega_{\mathbf{k}_1}-\omega_{\mathbf{k}_2});\nonumber\\
R^{S\leftrightarrow A+A}_{\mathbf{k},\mathbf{k}_1,\mathbf{k}_2}=W^{S\rightarrow A+A}_{\mathbf{k},\mathbf{k}_1,\mathbf{k}_2}\left(N_\mathbf{k}n_{\mathbf{k}_1}+N_\mathbf{k}n_{\mathbf{k}_2}\right.\\\left.-n_{\mathbf{k}_1}n_{\mathbf{k}_2}\,\right)\delta(\mathbf{k}-\mathbf{k}_1-\mathbf{k}_2)\delta(\omega_\mathbf{k}-\omega_{\mathbf{k}_1}-\omega_{\mathbf{k}_2})\nonumber.
\label{R_StoA} \end{gather}
The interaction probability is expressed via the amplitudes of the processes by the golden rule:
$W^{S\rightarrow A+A}_{\mathbf{k},\mathbf{k}_1,\mathbf{k}_2}=2\pi\vert V^{S\rightarrow A+A}_{\mathbf{k},\mathbf{k}_1,\mathbf{k}_2}\vert^2$, and analogously for $W^{S\rightarrow A+A}_{\mathbf{k},\mathbf{k}_1,\mathbf{k}_2}$. 
The corresponding amplitudes were calculated by \citet{Lyubarsky19}:
\begin{gather}
V^{S\rightarrow A+S}_{\mathbf{k}\mathbf{k}_1\mathbf{k}_2}=i\sqrt{\frac{\pi\omega_1}{2\omega\omega_2}}
\frac{k_{1\perp}\left[k_2-\sgn(k_{1\|})k_{2\|}\right]}{B_0k_{\perp}k_{2\perp}}\left(\mathbf{\hat{z}\cdot k}_1\times\mathbf{k}_2\right);\label{V_S to S+A}\\
V^{S\rightarrow A+A}_{\mathbf{k}\mathbf{k}_1\mathbf{k}_2}=i\sqrt{\frac{2\pi\omega_1\omega_2}{\omega}}\left(\frac{k_{1\perp}(\mathbf{k}_{2\perp}\cdot\mathbf{k}_{\perp})}{k_{2\perp}}\right.\nonumber\\\left.+\frac{k_{2\perp}(\mathbf{k}_{1\perp}\cdot\mathbf{k}_{\perp})}{k_{1\perp}}\right)\frac {H(-k_{1\|}k_{2\|})}{B_0k_{\perp}}.\label{V_S to A+A}
\end{gather}
Here the indexes $\|$ and $\perp$ describe the components of the $\mathbf k$ vector parallel and perpendicular to the background magnetic field.
In equation (\ref{V_S to S+A}), the index 1 is for the Alfv\'en wave and 2 for the fms wave.
In equation (\ref{V_S to A+A}), the Heaviside step function, $H(x)$, expresses the well-known fact that Alfv\'en waves do not interact if they propagate in the same direction along the magnetic field. 

\section{Decay of fms waves; qualitative considerations}

We study the possible decay of fms radiation produced by a strong enough source in a highly magnetized medium. Let us consider for simplicity the time evolution of a spatially homogeneous, isotropic fms radiation with the characteristic frequency of the order of $\omega_0=ck_0$. Assume that the spectrum is moderately wide, $\Delta\omega\sim\omega$, and the total radiation energy is 
\begin{equation}
    U=\int \omega N_{\mathbf{k}}d^3\mathbf{k}.
\label{N0}\end{equation}
The fms waves decay into fms and Alf\'ven waves of smaller frequency. The induced decay is possible only into states which are not empty; therefore, we assume that a weak background of fms and Alf\'ven waves is present in the whole phase space.

The initial pulse decays into waves satisfying the conservation laws (\ref{conservation}); therefore, the occupation numbers in the new states grow until the reverse merging process balances the decay. This happens when the occupation numbers in all three states become comparable. The decay $S\to S+A$ occurs into states with $k_1$ and $k_2$ comparable with $k_0$, and each initial fms quantum could produce only one quantum in the state $\mathbf{k}_1$ and one quantum in the state $\mathbf{k}_2$. Therefore the equilibrium is achieved, and the decay stops after the energy of the initial peak decreases roughly two times. 

The important point is that the phase volume available for decay 
$S\to A+A$ is in fact infinite because the Alf\'ven waves could have the perpendicular component of their wave vector arbitrarily large. 
To demonstrate this, let us write the conservation laws (\ref{conservation}) with the account of the dispersion laws (\ref{fms_disp}) and (\ref{ALfven_disp}) explicitly:
\begin{gather}
    k=\vert k_{1\|}\vert+\vert k_{2\|}\vert;\\
    k_{\|}=k_{1\|}+ k_{2\|};\\
    \mathbf{k}_{\perp}=\mathbf{k}_{1\perp}+\mathbf{k}_{2\perp}.
\end{gather}
One sees that for a given $\mathbf{k}$, one finds $k_{1\|}$ and $k_{2\|}$ from the first two equations, whereas the third equation is satisfied with an arbitrary  $\mathbf{k}_{1\perp}$ by choosing $\mathbf{k}_{2\perp}=\mathbf{k}_{\perp}-\mathbf{k}_{1\perp}$. In particular, the fms wave could decay into two Alf\'ven waves with arbitrarily large but nearly oppositely directed perpendicular components of the wave vectors.
This implies that the phase volume available for the decay of an fms wave into a pair of Alfv\'en waves is infinite, so the fms pulse could decay significantly, practically to the background level. 

Substituting $\mathbf{k}_{2\perp}=\mathbf{k}_{\perp}-\mathbf{k}_{1\perp}$ into equation (\ref{V_S to A+A}) and expanding in small $k/k_{1\perp}$, one finds the interaction amplitude in the limit $k_{1\perp}\gg k$: 
\begin{equation}
 V^{S\rightarrow A+A}_{\mathbf{k}\mathbf{k}_1\mathbf{k}_2}=i\sqrt{\frac{2\pi\omega_1\omega_2}{\omega}}k_{\perp}\left(1-\frac{(\mathbf{k}_{\perp}\cdot \mathbf{k}_{1\perp})^2}{k_{\perp}^2k_{1\perp}^2}\right)\frac {H(-k_{1\|}k_{2\|})}{B_0}.   
\label{V_StoA1}\end{equation}
One sees that the rate of the fms decay into Alfv\'en waves with large perpendicular components of the wave vector does not go to zero.

The population of such states is described by equation (\ref{kinetic_Alfven}), in which only the second term in the rhs should be retained. Taking into account that initially the occupation numbers for these states are small, one can write (note that in this equation, $\mathbf{k}_2$ is referred to fms waves whereas $\mathbf{k}$ to Alfv\'en waves)
\begin{equation}
\frac{\partial n_\mathbf{k}}{\partial t}=\sum_{\mathbf{k}_2}W^{S\rightarrow A+A}_{\mathbf{k}_2,\mathbf{k},\mathbf{k}_2-\mathbf{k}}N_{\mathbf{k}_2}\left(n_{\mathbf{k}}+n_{\mathbf{k}_2-\mathbf{k}}\right)\delta(\omega_{\mathbf{k}_2}-\omega_{\mathbf{k}}-\omega_{\mathbf{k}_2-{\mathbf{k}}})\end{equation}
Here we could take   $n_{\mathbf{k}}\approx n_{\mathbf{k}_2-\mathbf{k}}$ because adding a quantum into one of these states is accompanied by adding a quantum into another state. Then one finally finds that population of the Alfv\'en waves with $k_{\perp}\gg k_0$ grows initially (as soon as $n_\mathbf{k}\ll N_\mathbf{k_0}$) exponentially,
\begin{equation}
    \frac{\partial n_\mathbf{k}}{\partial t}= qn_\mathbf{k},
\end{equation}
with the rate
\begin{gather}
q=2\int W^{S\rightarrow A+A}_{\mathbf{k}_2,\mathbf{k},\mathbf{k}_2-\mathbf{k}}N_{\mathbf{k}_2}\delta(\omega_{\mathbf{k}_2}-\omega_{\mathbf{k}}-\omega_{\mathbf{k}_2-{\mathbf{k}}})d\mathbf{k}_2\\
=\frac{6\pi^3c}{B_0^2} \int\frac{\vert k_{\|}\vert (k_2-\vert k_{\|}\vert)k_{2\perp}^3}{k_2} N_{\mathbf{k}_2}H\left(k_\|(k_\|-k_{2\|})\right)\nonumber\\\times\delta(k_2-\vert k_{\|}\vert-\vert k_{2\|}-k_{\|}\vert)dk_{2\|}dk_{2\perp}
\end{gather}
Here we used equation (\ref{V_StoA1}) with an appropriate permutation of indexes $\mathbf{k}$, $\mathbf{k}_1$ and $\mathbf{k}$. 
Taking into account that  $k_{\|}\sim k_{2\|}\sim k_{2\perp}\sim k_2=\omega/c$ (according to the conservation laws, the longitudinal components of the wave vectors of the Alfv\'en and fms waves are comparable, whereas all components of the fms wave vector are typically of the order of $\omega/c$), one can find a rough estimate: 
\begin{equation}
    q\sim q_0=\frac{8\pi}{B_0^2} U\omega.
\label{q0}\end{equation}
The coefficient in the definition of $q_0$ is chosen to show explicitly that the interaction rate is proportional to the small ratio of the wave energy and the energy of the background magnetic field. 

The total number of the produced Alfv\'en quanta, ${\cal N}=\int n_\mathbf{k}d\mathbf{k}$, grows as
\begin{equation}
    \frac{\partial\cal N}{\partial t}=2\pi\int qn_\mathbf{k}k_{\perp}dk_\perp dk_\|.
\end{equation}
Here the integration over $k_\|$ is limited, due to the conservation laws, by the region $k_\|\sim k_0$, whereas the integral over $k_\perp$ is unlimited. 
Taking into account that  $q$ is independent of $k_{\perp}$ at large $k_{\perp}$, one sees that the total production rate of the Alfv\'en quanta diverges unless $n_\mathbf{k}$ decreases with $k_\perp$ faster than $k_{\perp}^{-2}$. Of course, Alfv\'en waves are erased by dissipation processes at small enough wavelengths, but the corresponding $k$ is typically too high so that the Alfv\'en waves would be produced at an inappropriately high rate unless the background spectrum is steep enough. Here we adopt a natural assumption that the background waves have the spectrum $n_\mathbf{k}\propto k^{-\alpha}$ with $\alpha>2$.

In this case, the evolution of the system could be qualitatively described as follows. At the initial stage, the background Alfv\'en spectrum exponentially grows with the rate of eq.\ (\ref{q0}) at all $k_\perp$. The waves with $k_\perp\sim k_0$ reach the saturation level, $n_{\mathbf{k}}\sim N_{\mathbf{k}}$, first. At this stage, the fms pulse weakens a few times. The fms waves keep decaying into Alfv\'en waves with larger $k_\perp$ so that the energy of fms waves decreases monotonically. Then the Alfv\'en waves with smaller $k_\perp$ begin to merge into fms wave to maintain the equilibrium $n_{\mathbf{k}}\sim N_{\mathbf{k}}$. Thus the Alfv\'en population of the states with larger $k_\perp$ gradually increases until the saturation level, $n_{\mathbf{k}}\sim N_{\mathbf{k}}$. At smaller $k_\perp$, a plateau is formed, which gradually decreases with time. 
In such a way, the fms pulse eventually could decay to the background level. The important point is that the decay rate is determined by the instant energy of the fms pulse. Since the energy of fms waves decreases, the decay process slows down with time. Therefore in real systems, the energy of fms waves decays until the decay rate becomes equal to the wave escape rate. Therefore the process places limits on the luminosity of the source. In the next section, we present numerical simulations that confirm the picture described.

\section{Simulations}

We study the time evolution of a spatially homogeneous, isotropic system of waves described by the kinetic equations (\ref{kinetic_fms}) and (\ref{kinetic_Alfven}).
We assume that the initial spectral distribution of fms waves  has a bell-shaped form,
\begin{equation}
N_{\mathbf{k}}(t=0)=N_0\exp\left[-\left(\frac{k-k_0}{k_0}\right)^2\right].
\end{equation}
In addition, we assume that the whole phase space is filled by the background  fms and Alv\'en waves 
with the spectrum
\begin{equation}
N_{\mathbf{k}}^b=n_{\mathbf{k}}^b=10^{-4}N_0\left\{\begin{array}{ll}1;
\qquad k<0.5k_0;\\
(0.5+k/k_0)^{-\kappa};\qquad  k\geq0.5k_0.\end{array}\right.
\end{equation}
In simulations, we used the power law index $\kappa$  from $2.5$ to $3.5$ and found that the results differ insignificantly. Below, we present results  for $\kappa=3$.

We solve the system of equations (\ref{kinetic_fms}) and (\ref{kinetic_Alfven})  using the Runge-Kutta method of the second order with correction (see Appendix). The system is closed; therefore, the total energy of the system,  $U_0 = \int \omega(N_{\mathbf k}+n_{\mathbf k})d^3k$, is conserved. 
\begin{figure}[!h]
{
	\centering
	\includegraphics[width=0.53\textwidth]{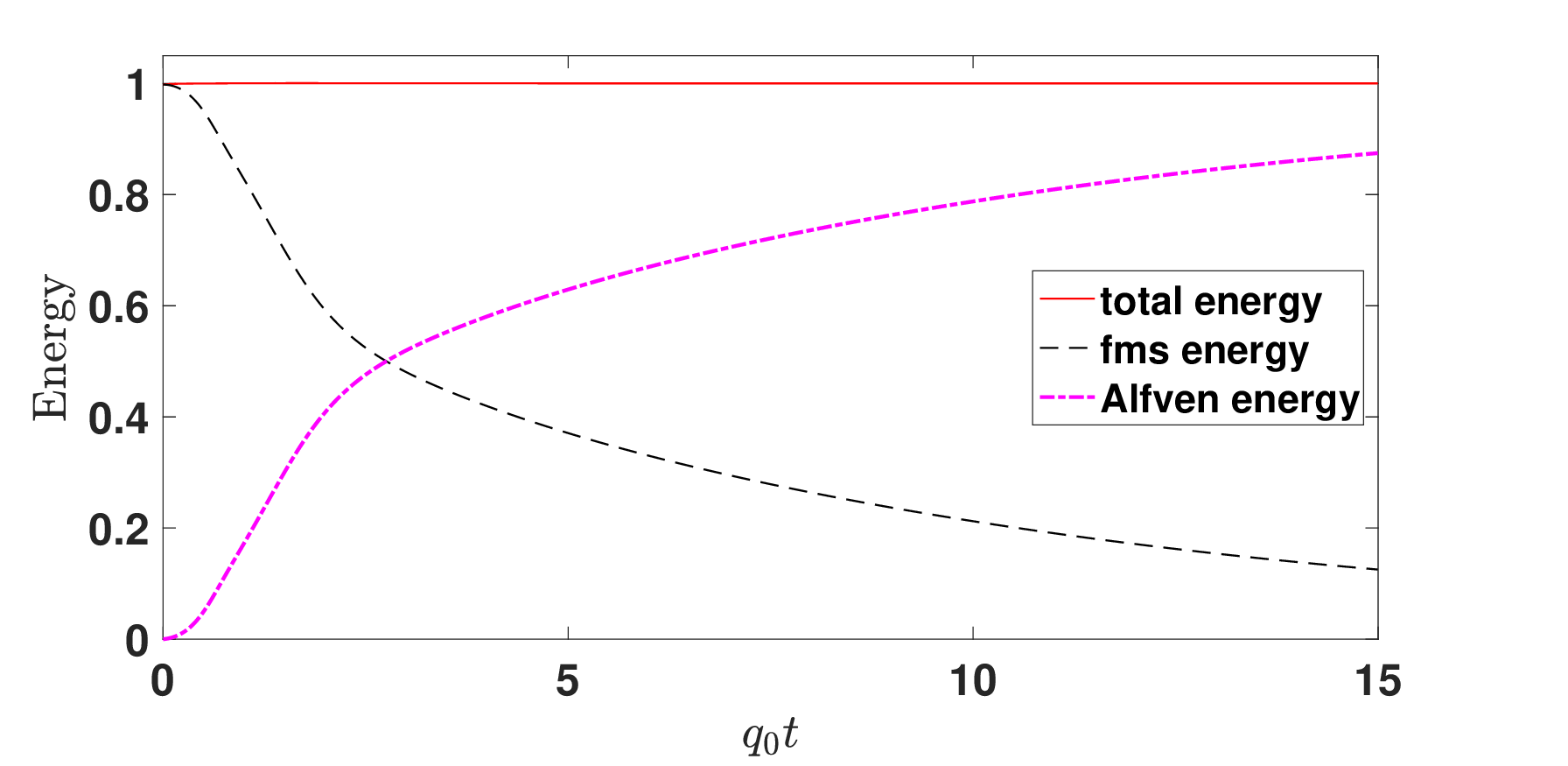}
	\caption{Time evolution of the energy of fms and Alfv\'en waves  normalized to the total energy.}.}\label{Energy}
\end{figure}
Figure 1 shows the time evolution of the energies of fms and Alfv\'en waves normalized to the system's total energy, which is chosen to be equal to unity. One sees that the fms waves are efficiently converted to the Alfv\'en waves with the rate given by equation (\ref{q0}). 

The evolution of spectra is shown in figs.\ 2 and 3.  We plotted the spectral distributions integrated over the longitudinal wave vectors because the evolution of the perpendicular component of the wave vector is the most interesting. 
The presented results confirm the qualitative picture 
described in the previous section.
Namely, the population of the fms gradually decreases. The population of the Alfv\'en waves grows across the whole spectrum until both populations become comparable, $n_{\mathbf{k}}\sim N_{\mathbf{k}}$, at $k_\perp\sim k_0$. 
This occurs at $q_0t\sim 1-2$. 
At larger times, only the Alfv\'en population of the high $k_\perp$ states grows, whereas at smaller $k_\perp$,  a plateau is formed, the level of which slowly decreases.

\begin{figure}[!h]
{
	\centering
	\includegraphics[width=0.54\textwidth]{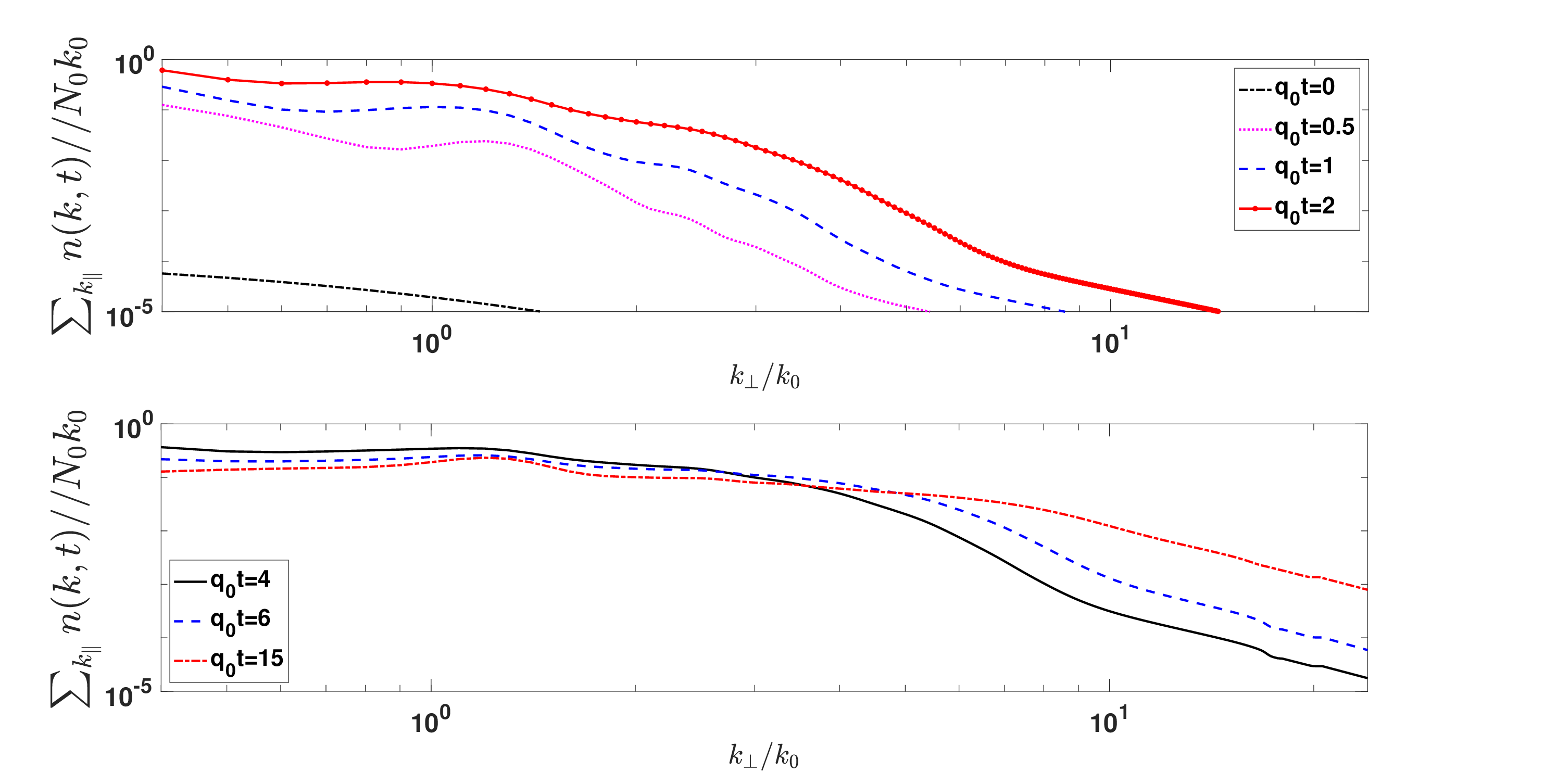}
	\caption{The time evolution of spectra of Alfv\'en waves integrated in   $k_{\Vert}$. 
	}.}\label{As}
\end{figure}

\begin{figure}[!h]
{
	\centering
	\includegraphics[width=0.53\textwidth]{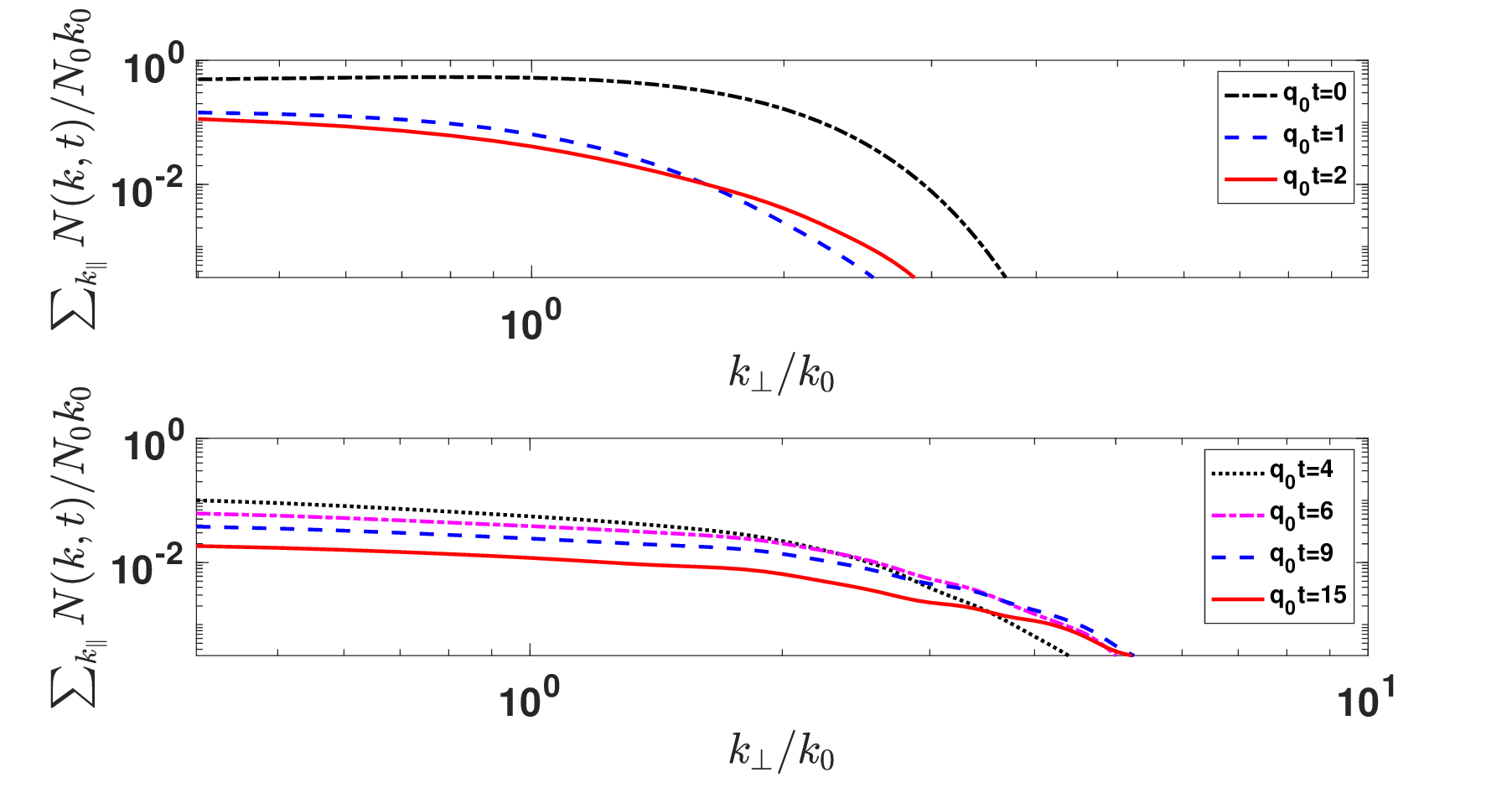}
	\caption{The time evolution of spectra of fms waves integrated in $k_{\Vert}$. 
	}.}\label{As}
\end{figure}

\section{ASTROPHYSICAL APPLICATIONS}

The nonlinear interactions of low-frequency waves play an important role in compact sources of powerful radio emission, such as pulsars and FRBs. The results of this paper could be applied if the waves may be described in the MHD limit, i.e., when the emission frequency is well below both the Larmor and the plasma frequencies. 

\subsection{Fast radio bursts}
FRBs are radio pulses of millisecond duration coming from cosmological distances and having isotropic luminosities $L_{\rm iso}\sim 10^{42}-10^{45}$ erg$\cdot$s$^{-1}$. The origin of these pulses is still not known; however, there is some evidence that they are associated with magnetar flares (see, e.g., review by \citealt{Zhang22} and references therein). In the magnetar magnetospheres, both the Larmor and the plasma frequencies are well above the radio band; therefore, independently of the emission mechanism, radio waves propagate as fms waves. Then, the nonlinear decay of fms into the Alfv\'en waves places severe limits on radio emission power if FRBs are produced well within the magnetar magnetospheres.

The transformation efficiency is determined by 
the product of the transformation rate (\ref{q0}) and the propagation time, $r/c$. Assuming the dipole magnetic field in the magnetosphere, $B=\mu/r^3$, and expressing the fms energy density via the isotropic FRB luminosity, $U=L_{\rm FRB}/4\pi cr^2$, one finds 
\begin{equation}
    \frac{q_0r}c=\frac{2L_{\rm FRB}\omega r^5}{\mu^2 c^2}=13\frac{f_9L_{\rm FRB, 43}r_7^5}{\mu^2_{33}}.
\end{equation}
Here $f=\omega/2\pi$ is the radiation frequency, and we employ the standard short-hand notation,  $s=10^xs_x$ in cgs units. One sees that at any FRB power, the radiation is absorbed at a distance from a few to a few dozen stellar radii.

This estimate assumes that the turbulence is weak, i.e., the ratio of the wave energy to the energy of the background field is small. A simple estimate shows that this condition is fulfilled in the region of interest:
\begin{equation}
    \frac{8\pi U}{B^2}=7\cdot 10^{-6}\frac{L_{\rm FRB,43}r_7^4}{\mu^2_{33}}.
\end{equation}
This justifies the neglect of four-wave processes\footnote{ Note that the rate of the four-wave interaction, $q\sim (8\pi U/B_0)^2\omega$, remains larger than the escape rate, $c/r$, for $L_{\rm FRB}>10^{37}$ erg$\cdot$ s$^{-1}$. Therefore cascading of the excited Alfv\'en waves may be possible. However, the rate of three-wave processes is larger; therefore, we take into account only fms-to-Alfv\'en interaction.
 }.

Note that the wave amplitude exceeds the background field when this ratio exceeds unity. In this case, the fms radiation is heavily absorbed because the MHD condition, $E<B$, is violated \citep{Beloborodov21,Beloborodov23}. If the background field is a dipole, this happens at distances of a few hundred stellar radii. However, the magnetic disturbance from the magnetar flare propagates away as a large-scale electro-magnetic pulse, whose amplitude decreases as $r^{-1}$. In this case, the high-frequency waves propagate at the top of the pulse. Then, the wave amplitude remains smaller than the background field, so Beloborodov's mechanism does not work. On the other hand, the fms-to-Alfv\'en transformation provides an effective absorption in any case.

The important point is that the wave transformation is a stimulated process, i.e., the transformation rate is proportional to the occupation number of the waves in the final state. The high rate is obtained because the initially small density of Alfv\'en waves grows exponentially and rapidly becomes comparable with the density of fms waves. The transformation could be suppressed if an efficient absorption mechanism does not permit the growth of Alfv\'en waves.
The newly produced Alfv\'en waves could be absorbed because of the current starvation, i.e., when the parameter
\begin{equation}
    \xi=\frac j{enc}
\label{xi}\end{equation}
exceeds unity \citep{Thompson_Gill14,Thompson23}. Here $n$ is the plasma density, $j=ck_\perp \delta B/4\pi$ the current density in the Alfv\'en wave with the amplitude $\delta B$. Let us consider this process. 

In the magnetosphere of an active magnetar, the electron–positron pairs are produced by slow untwisting of magnetospheric magnetic field lines \citep{BeloborodovThompson07,Beloborodov13a}. The plasma density is estimated as \citep{Beloborodov20} 
\begin{equation}
    n=\frac{{\cal M}\mu}{4\pi er^3r_{\pm}},
\end{equation}
where ${\cal M}\sim 10^3$ is the pair multiplicity, $r_\pm= 5\cdot 10^6 \mu_{33}^{1/3}$ cm the distance from the star where the magnetic field
falls to $10^{13}$ G, so the pair production stops. Substituting this estimate into eq.\ (\ref{xi}) and  assuming that the FRB energy is completely transferred to Alfv\'en waves, $L_{\rm FRB}=\delta B^2r^2c$, one finds
\begin{equation}
    \xi=\frac{k_\perp r^2r_\pm}{{\cal M}\mu}\sqrt{\frac{L_{\rm FRB}}c}=1.9\frac{r^2_7f_9L^{1/2}_{\rm FRB,43}}{{\cal M}_3\mu_{33}^{2/3}}\frac{ck_\perp}{\omega}.
\end{equation}
One sees that the current starvation sets in only when a significant fraction of the FRB energy is transferred to Alfv\'en waves. In this case, the FRB is absorbed via the transformation to Alfv\'en waves, which decay because of current starvation. However, the important point is that the pairs are heated to high Lorentz factors and produce new pairs. 

Let us assume that a fraction $\zeta<1$ of the FRB energy is absorbed. Then the acquired Lorentz factor is estimated as
\begin{equation}
    \gamma=\frac{\zeta L}{4\pi r^2m_ec^3n}=
 10^7\,\frac{r_7\zeta L_{\rm FRB, 43}}{{\cal M}_3\mu_{33}^{2/3}}.
\end{equation}
The particles with this Lorentz factor emit curvature photons with the energy
\begin{equation}
    \varepsilon=\frac{\hbar c}r\gamma^3=2\cdot 10^3\frac{(\zeta L_{\rm FRB, 43})^{3}r_7^{2}}{{\cal M}_3^{3}\mu_{33}^{2}}\,\rm MeV.
\end{equation}
These photons produce pairs just as in pulsars. 
The condition for single photon pair production is
\begin{equation}
    \chi=\frac{\varepsilon\sin\theta B}{2m_ec^2 B_q}>0.1.
\label{chi}\end{equation}
Here $\theta$ is the angle between the photon direction and the magnetic field, $B_q=m_e^2c^3/\hbar e^2=4.4\cdot 10^{13} G$ the quantum magnetic field. The photon is emitted along the magnetic field and, after passing the distance $x$, acquires the angle $\theta=x/r$. Now one finds
\begin{equation}
   \chi=45\frac xr \frac{(\zeta L_{\rm FRB, 43})^{3}}{{\cal M}_3^{3}\mu_{33}r_7}.
\end{equation}
The condition (\ref{chi}) is fulfilled, so all the emitted photons are converted to pairs. One can easily check that the power of the curvature emission is sufficient for the particle to lose the whole energy to radiation. Therefore each particle emits
\begin{equation}
    {\cal N}=\frac{m_ec^2\gamma}{\varepsilon }=2.5\cdot 10^3\frac{{\cal M}_3^2\mu_{33}^{4/3}}{(\zeta L_{\rm FRB, 43})^{2}r_7}.
\end{equation}
More pairs are produced from the synchrotron photons emitted by the newly produced pairs. In any case, one sees that absorption of a fraction of the FRB energy produces enough pairs to provide conditions for fms-to-Alfv\'en decay in the MHD regime.


The above consideration shows that FRBs could not be generated well within the magnetar magnetosphere. However, one could not directly extrapolate the obtained conclusion to the outer magnetosphere or the magnetar wind. First of all, the plasma density rapidly decreases with the distance, so eventually, waves in the radio band could not be described in the MHD approximation. Moreover, the magnetic perturbation from the magnetar flare propagates away as a large-scale MHD pulse, which amplitude decreases as $1/r$, so that in the outer magnetosphere, the magnetic field of the pulse exceeds the dipole field and the plasma is pushed away with relativistic velocities. Therefore the ratio of the radiation energy density to the energy density of the background field stops growing, and moreover, one has to take into account the relativistic slowing down of time. This implies that magnetar flares could produce FRBs only far enough from the magnetar (see, e.g., review by \citealt{Lyubarsky21} and references therein).

\subsection{Radio emission of the Crab pulsar}

A typical pulsar produces a pencil beam of radio emission, presumably generated in the electron-positron plasma flowing along the magnetic axis of the neutron star within a narrow open field line tube. The emission is generally attributed to plasma oscillations in the flow (see the recent review by \citealt{Philippov_Kramer22} and references therein). The frequency of these waves is comparable with the plasma frequency in the plasma rest frame; therefore, they could not be considered MHD waves. The nonlinear process under consideration is irrelevant to this emission. 

In pulsars with large magnetic fields at the light cylinder, such as the Crab and millisecond pulsars, there is another emission site, namely, the current sheet separating, beyond the light cylinder, the oppositely directed magnetic fields. The energy release due to the magnetic reconnection in the current sheet feeds the powerful synchrotron emission in the gamma-ray, and sometimes also in the X-ray and optical, band \citep{Lyubarskii96,Bai_spitkovsky10,Cerutti_etal16}. The fan beam thus formed rotates with the neutron star so that the observer typically sees two peaks per pulsar period. Some of these pulsars also exhibit radio pulses in phase with high-energy pulses. This radio emission could be produced because  magnetic islands in the reconnecting current sheet continuously merge, giving rise to magnetic perturbations that propagate away in the form of fms waves, which further away are transformed into radio waves \citep{Uzdensky_Spitkovsky14,Lyubarsky19,Philippov19}. The nonlinear interaction of fms waves places limits on the radio luminosity of these pulsars.

According to simulations by \citet{Philippov19}, about 0.5\% of the total energy release in the current sheet is radiated away in the form of low-frequency waves. The luminosity of the Crab pulsar in the X- and $\gamma$-ray bands is roughly $L_{\rm hard}=10^{36}$ erg$\cdot$s$^{-1}$. This quantity could be considered a proxy for the energy release rate in the current sheet. Then, one would expect  the radio luminosity of the Crab to be of the order of $L'=5\cdot 10^{33}$ erg$\cdot$s$^{-1}$. However, the observed radio luminosity is two orders of magnitude smaller, $L_{\rm radio}=7\cdot 10^{31}$ erg$\cdot$s$^{-1}$ (e.g., \citealt{Malov_Malofeev94}). This may be attributed to the decay of the fms into the Alfv\'en waves on the way out of the magnetosphere.

The decay rate (\ref{q0}) is calculated in the zero electric frame of the plasma because the non-linear interactions in the force-free regime are not affected by plasma
moving along the magnetic field lines. Just beyond the light cylinder, the magnetospheric electric and magnetic fields are of the
same order but not too close to each other so that the velocity of the zero electric field frame, $\mathbf{v}=c\mathbf{E\times B}/B^2$, is only mildly relativistic. Therefore we use the parameters in the lab frame. We find the radiation energy density, $U$, from the condition that the decay rate (\ref{q0}) is comparable with the wave escape rate, $q_0\sim c/r$. The radiation is produced near the light cylinder; therefore, $r\sim c/\Omega$, where $\Omega=2\pi/P$ is the angular velocity of the neutron star, and $P$ is the pulsar period. Now this condition yields
\begin{equation}
\frac{8\pi U}{B^2}\sim\frac{\Omega}{\omega},
\end{equation}
The magnetic field in the equatorial zone at the distance of the light cylinder is $B=2\mu(\Omega/c)^3$, where $\mu$ is the magnetic moment of the neutron star. The last is related to the pulsar spin-down power:
\begin{equation}
    L_{\rm sd}= (1+\sin^2\psi)\frac{\mu^2\Omega^4}{c^3},
\end{equation}
 where $\psi$ is the angle between the magnetic and rotational axes \citep{Spitkovsky06}.
The radio emission forms a fan beam with the opening angle $\alpha\sim 0.1$, so that the radio luminosity may be presented as
\begin{equation}
    L_{\rm radio}= 2\pi\alpha Uc(c/\Omega)^2.
\label{Lradio}\end{equation}
 Now one finds
\begin{equation}
    \frac{L_{\rm radio}}{L_{\rm sd}}\sim \frac{\alpha}{(1+\sin^2\psi)Pf},
\end{equation}
where $f=\omega/2\pi$ is the radiation frequency. The period of the Crab pulsar is $P=0.033$ s, and the spectrum is very steep without the low-frequency cutoff down to the decameter band. Therefore, we take a low frequency $f=30$ MHz \citep{Malov_Malofeev94}. Then one gets $L_{\rm radio}/L_{\rm sd}\sim 10^{-7}$. Taking into account that the observed slowing down rate of the Crab corresponds to the spin-down power $L_{\rm sd}=5\cdot 10^{38}$ erg$\cdot$s$^{-1}$, one sees that the obtained estimate is compatible with the observed radio luminosity.

Now let us check that the MHD conditions are fulfilled, i.e., the plasma density is sufficient to maintain Alfv\'en waves. The plasma density in pulsars may be presented as
\begin{equation}
n=\kappa\frac{\Omega B}{2\pi e},
\end{equation}
where $\kappa$ is the multiplicity. In young pulsars, $\kappa>10^5$ \citep{Timokhin_Harding15}, which is compatible with the observations of the Crab Nebula \citep{deYager96}. Assume that the emitted power, $L'$, (which is larger than the observed luminosity, $L_{\rm radio}$, see above) is transformed into Alfv\'en waves so that the relation between $L'$ and the Alfv\'en energy density is the same as the relation (\ref{Lradio}) between the observed luminosity and the radiation energy density.  Now we estimate the parameter $\xi$ (see eq.\ \ref{xi}) as
\begin{align}
    \xi=\frac{fP}{2\alpha\kappa}\sqrt{\frac{L'}{2(1+\sin^2\psi)L_{\rm sd}}}\frac{k_\perp c}{\omega}\\=0.2\frac{f_{7.5}P_{-1.5}L'^{1/2}_{33.7}}{\alpha_{-1}\kappa_5(1+\sin^2\psi)L_{\rm sd,38.7}^{1/2}}\frac{k_\perp c}{\omega}.\nonumber
\end{align}
 One sees that the density is marginally sufficient to maintain fms-to-Alfv\'en transformation at the expected luminosity of the current sheet. The produced Alfv\'en waves decay because of current starvation when the energy is transferred to higher $k_\perp$, as is described in previous sections.
 This justifies our explanation of the low luminosity of the Crab pulsar.

\section{Conclusions}

In this paper, we addressed the  nonlinear decay of fms waves in relativistic force-free MHD. There are models of pulsars and FRBs, in which the radio emission is generated in dense magnetospheres such that not only the Larmor but also the plasma frequency is well below the radiation frequencies (see, e.g., reviews by \citealt{Zhang22,Lyubarsky21,Philippov_Kramer22}). Within these sources, the radiation propagates in the form of fms waves, and the nonlinear decay of fms into Alfv\'en waves 
could strongly affect the properties of the outgoing radiation. 

Using the kinetic equations for the waves, we estimated the decay rate and studied the kinetics of the decay process. We have shown that an fms wave could decay into two Alfv\'en waves with arbitrary large wave vectors if these wave vectors are nearly perpendicular to the background magnetic field and nearly oppositely directed. Therefore the phase volume available for the decay of an fms wave is in fact infinite. In this case, the energy of fms waves could be completely transferred to the small-scale Alfv\'en waves not via a cascade, as in the Kolmogorov turbulence, but directly. Numerical solutions of the kinetic equations confirmed these conclusions.
Our results explain the anomalously low radio efficiency (the ratio of the radio to the spin-down power) of the Crab pulsar and demonstrate that FRBs could not be produced well within magnetar magnetospheres. 

\section*{Acknowledgments}

 We are grateful to the anonymous referee for insightful comments.
This research was supported by grant I-1362-303.7/2016 from the German-Israeli Foundation for
Scientific Research and Development and by grant 2067/19 from the Israeli Science Foundation.
\section*{Appendix. Numerical procedure and convergence.}
The set of equations (\ref{kinetic_fms}) and (\ref{kinetic_Alfven})  was solved by the Runge-Kutta method. It is known that the determination of the error and stability of the Runge-Kutta method is quite difficult \cite{Butcher15}. As a rule, a very small time step is necessary to minimize the error and ensure the stability of the method. However, if there are conserved parameters in the system (e.g., total energy, total mass, helicity, etc.), one can achieve better convergency and stability by making use of a correction procedure that explicitly exploits the conservation laws (see, e.g.,  \citealt{Christlieb,Palha,Coppola} and references there). 
 
 In our case, the conserved parameter is the total energy of the system, 
 \begin{equation}
 U = \int \omega(N_{\mathbf k}+n_{\mathbf k})d^3k.
 \end{equation}
This means that 
summing up the rhs of equations (4) and (5), multiplying the obtained expression by $\omega$ and integrating over the whole phase space yields zero.  However, the conservation law is violated when the integrals in the rhs of these equations are evaluated numerically. To keep the total energy exactly conserved, we introduce a  correction to the obtained values of $N(\mathbf k)$ and $n(\mathbf k)$ at each time step.
 
Let $U_1$ be the energy of the system obtained after the $i$-th time step, and $U_0$ be the initial energy of the system. We multiply $N_{i+1}(\mathbf k)$ and $n_{i+1}(\mathbf k)$ by $\xi=2-\frac{U_1}{U_0}$.  Taking into account the identity
\begin{equation}
U_1(2-\frac{U_1}{U_0})
=U_0(1-(1-\frac{U_1}{U_0})^2),
\end{equation}
one sees that after such a correction, the total energy of the system remains equal to $U_0$ to within $(1-\frac{U_1}{U_0})^2$.
In our  calculations 
the value of $1-\frac{U_1}{U_0}$ does not exceed $10^{-3}$. Consequently, the energy is conserved to within $\sim 10^{-6}$.

 \begin{figure}[!h]
{
	\centering
	\includegraphics[width=0.45\textwidth]{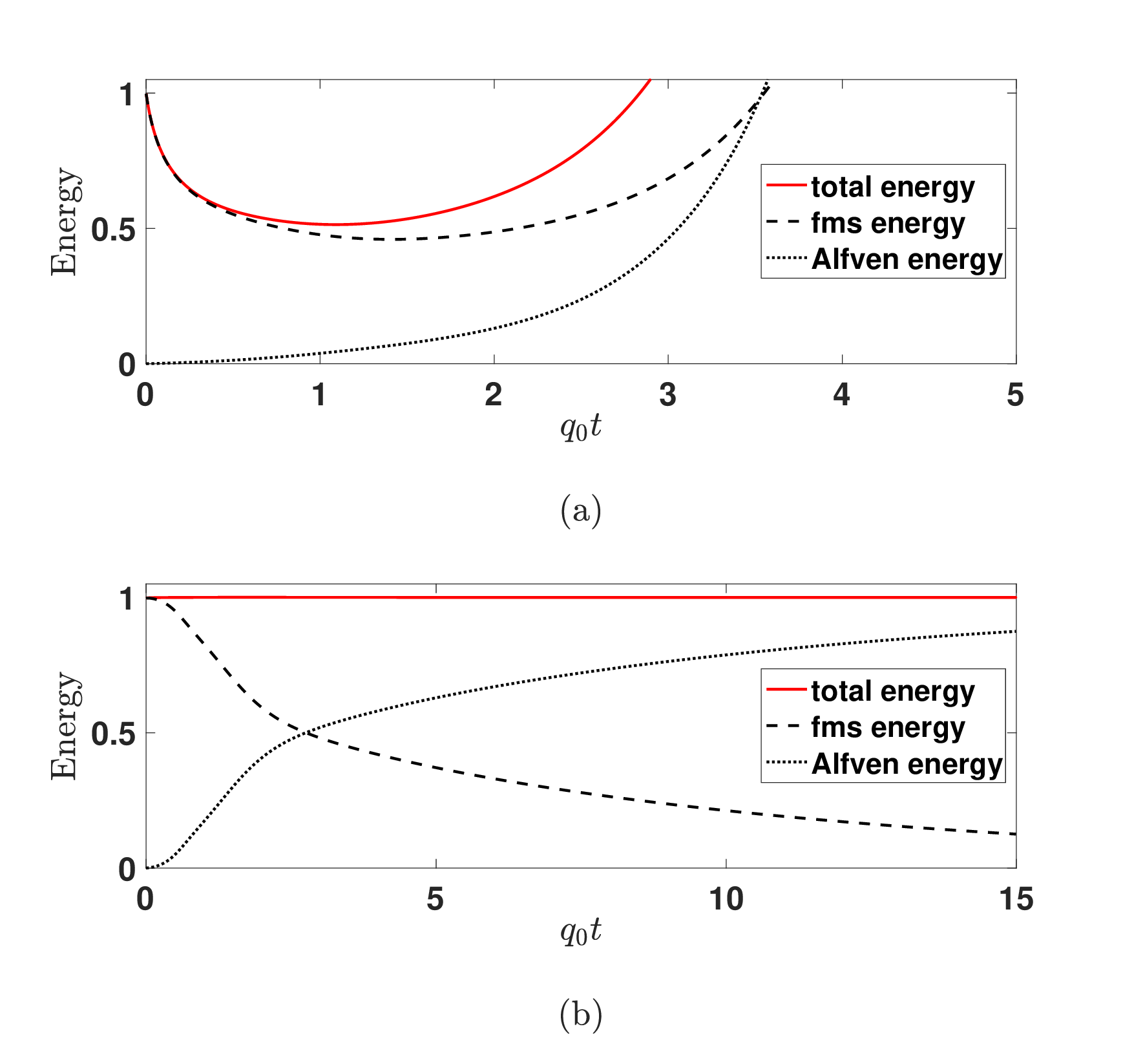}
	\caption{Distribution of normalized to the total energy of the system, energy fms and Alfv\'en waves in time. (a) without 
	the correction; (b) with the correction}}\label{Energy1}
\end{figure}

In figure 4, we show how the above correction procedure 
affects the stabilization of the calculations. The calculations were carried out with  the time step $q_0\Delta t=10^{-4}$ and values $\Delta k_{\perp}= \Delta k_{\parallel}= 0.1k_0$. One sees that without correction (Fig.4a), the system quickly becomes unstable. 
When using the correction procedure (Fig.4b), the system remains stable, and the total energy of the system is conserved.

 \begin{figure}[!h]
{
	\centering
	\includegraphics[width=0.47\textwidth]{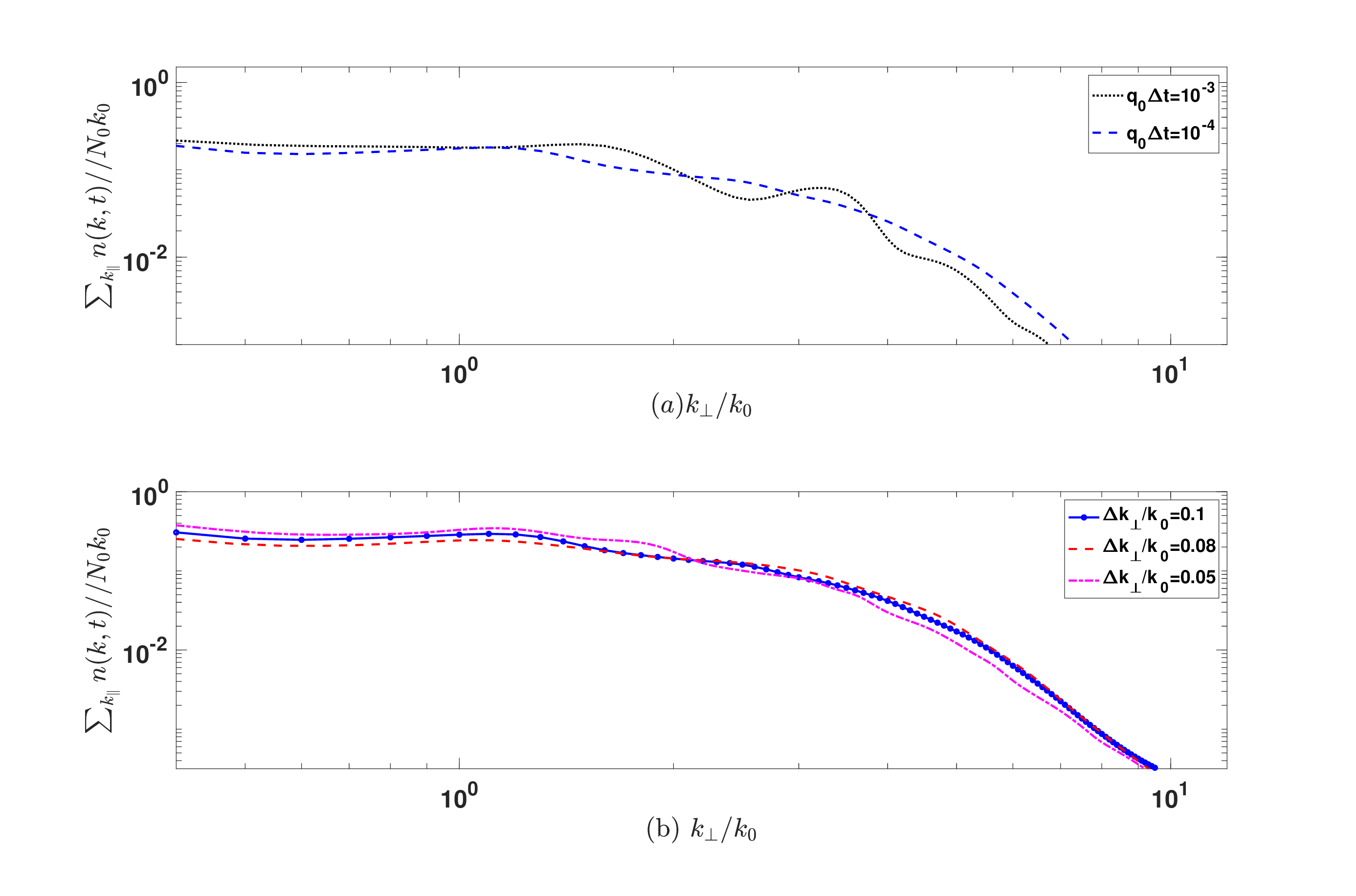}
	\caption{The spectrum of Alfv\'en waves at $q_0t=1$. (a) $\Delta  k_{\perp}/k_0=0.1$ and $q_0\Delta t =10^{-3}$ (black solid) and $10^{-4}$ (blue dashed.  (b) $q_0\Delta t =10^{-4}$ and $\Delta  k_{\perp}/k_0=0.1$ (blue dot dashed), $0.08$ (red dashed), $0.05$ (magenta dotted). }}\label{Var}
\end{figure}

Figure 5 shows the spectra of  Alfv\'en waves calculated with different input parameters. The spectra are presented at $q_0t=1$, when the energies of the fms and  Alfv\'en waves are comparable. 
Figure 5a shows the spectra calculated with different time steps, $q_0\Delta t$, at value $\Delta k_{\perp}/k_0=0.1$. One sees that when the time resolution is improved, the "waves" in the spectra disappear, whereas the overall spectral shape remains intact.  
Figure 5b shows the spectra obtained for different values $\Delta k_{\perp}/k_0$ at $q_0\Delta t=10^{-4}$. One sees that improving the resolution in the wave vector space does not significantly affect the shape of the spectrum. 

\bibliographystyle{apj}
\bibliography{FRB}

\label{lastpage}	

\end{document}